\documentclass[final,5p,times,twocolumn]{elsarticle}

\usepackage{amssymb}

\usepackage{physics, siunitx}
\RequirePackage[l2tabu, orthodox]{nag}
\usepackage[all, warning]{onlyamsmath}
\DeclareBinaryPrefix\kibi{Ki}{10}

\journal{Computer Physics Communications}

\begin{document}

\begin{frontmatter}



\title{Gravitational octree code performance evaluation on Volta GPU}
\author{Yohei Miki\fnref{label1}}
\fntext[label1]{Information Technology Center, The University of Tokyo, 5-1-5 Kashiwanoha, Kashiwa, Chiba 277-8589, Japan}
\ead{ymiki@cc.u-tokyo.ac.jp}




\begin{abstract}
In this study, the gravitational octree code originally optimized for the Fermi, Kepler, and Maxwell GPU architectures is adapted to the Volta architecture. 
The Volta architecture introduces independent thread scheduling requiring either the insertion of the explicit synchronizations at appropriate locations or the enforcement of the same implicit synchronizations as do the Pascal or earlier architectures by specifying \texttt{-gencode arch=compute\_60,code=sm\_70}. 
The performance measurements on Tesla V100, the current flagship GPU by NVIDIA, revealed that the $N$-body simulations of the Andromeda galaxy model with $2^{23} = \num{8388608}$ particles took \SI{3.8e-2}{s} or \SI{3.3e-2}{s} per step for each case. 
Tesla V100 achieves a \numrange{1.4}{2.2}-fold acceleration in comparison with Tesla P100, the flagship GPU in the previous generation. 
The observed speed-up of \num{2.2} is greater than \num{1.5}, which is the ratio of the theoretical peak performance of the two GPUs. 
The independence of the units for integer operations from those for floating-point number operations enables the overlapped execution of integer and floating-point number operations. 
It hides the execution time of the integer operations leading to the speed-up rate above the theoretical peak performance ratio. 
Tesla V100 can execute $N$-body simulation with up to $25 \times 2^{20} = \num{26214400}$ particles, and it took \SI{2.0e-1}{s} per step. 
It corresponds to \SI{3.5}{TFlop/s}, which is \num{22}\% of the single-precision theoretical peak performance. 
\end{abstract}


\end{frontmatter}


\section{Introduction}
\label{sec:introduction}
Tesla V100 is the leading graphics processing unit (GPU) based on the latest Volta architecture\footnote{The Turing architecture is the latest generation of GPU at the time of writing the manuscript, but its main focus is graphics. Besides, the compute capability of the Turing architecture is 7.5, which means that the major number is identical to that of the Volta architecture.} for high-performance computing by NVIDIA. 
The single-precision theoretical peak performance of Tesla V100 is \SI{15.7}{TFlop/s} \cite{GV100Whitepaper}, which is \num{1.5} times higher in comparison with Tesla P100, the previous flagship GPU. 
The primary driver of the performance improvement is the increase in the number of streaming multiprocessors (SMs) from \num{56} to \num{80}. 
One notable modification of SM in the Volta architecture is the independence of integer units from CUDA cores that were unified in the Pascal or earlier GPU architectures. 
The results of the detailed micro-benchmarks on Tesla V100 are reported by \cite{Jia2018}. 
For example, the measured load throughput of L1 cache per SM is more than three times higher than that of Tesla P100. 

Recently installed large-scale systems, such as Summit by Oak Ridge National Laboratory \cite{Summit} and ABCI (AI Bridging Cloud Infrastructure) by National Institute of Advanced Industrial Science and Technology \cite{ABCI}, employ Tesla V100 as the accelerator device. 
They appear at the top of the TOP500 and Green500 lists \cite{TOP500, Green500}. 
Such systems demand porting the software developed for the Pascal or earlier GPU architectures to the Volta architecture, including optimizing the software performance for it. 
Summarizing the pitfalls, preparing recipes for porting applications to the Volta architecture, and unveiling their performance behavior would help users to develop software for the Volta GPU architecture. 

This study focuses on $N$-body simulations as a representative example of compute-intensive applications. 
In astrophysics, $N$-body simulations are the common tool for investigating the formation and evolution processes of gravitational many-body systems such as galaxies. 
$N$-body simulations solve the time evolution of the system by tracking the orbital evolution of particles obeying the gravitational force between the particles. 
The acceleration of the $i$th particle out of $N$ particles is given by Newton's equation of motion as 
\begin{align}
  \vb*{a}_i &= \sum_{j = 0, j \neq i}^{N - 1}\frac{G m_j \pqty{\vb*{r}_j - \vb*{r}_i}}{\pqty{\abs{\vb*{r}_j - \vb*{r}_i}^2 + \epsilon^2}^{3 / 2}},
  \label{eq:eq.of.motion}
\end{align}
where $m_i$, $\vb*{r}_i$, and $G$ are the particle mass, position, and the gravitational constant, respectively. 
The gravitational softening $\epsilon$, which is introduced to reduce the artificial effects caused by the small particle number $N$ compared to the actual system, acts to remove the divergence due to division by zero and self-interaction. 

The direct method that calculates the gravitational force between particles one by one according to Equation~\ref{eq:eq.of.motion} is too slow to execute the simulations in a realistic time because the computation complexity is $\order{N^2}$. 
The tree method proposed by \cite{BarnesHut1986} reduces the workload to $\order{N \log N}$ by using a multipole expansion technique when calculating the gravitational force by distant particles. 
Using accelerator devices is another way of accelerating $N$-body simulations. 
Earlier implementations have shown that GPUs can accelerate the tree method \cite{Gaburov2010, Nakasato2012, Ogiya2013, Bedorf2012, Bedorf2014, WatanabeNakasato2014, MikiUmemura2017}. 

\texttt{GOTHIC}, the gravitational octree code developed earlier \cite{MikiUmemura2017}, adopts both the tree method and the block time step \cite{McMillan1986}. 
The code was initially developed for the Fermi GPU architecture using CUDA C/C++ and was optimized for the Fermi, Kepler, and Maxwell architectures (see \cite{MikiUmemura2017} for the implementation details). 
\texttt{GOTHIC} automatically adjusts the frequency of rebuilding the tree structure to minimize the time-to-solution by monitoring the execution time of the tree construction and the gravity calculation. 
\texttt{GOTHIC} exploits the acceleration MAC (multipole acceptance criterion) proposed by \cite{Springel2001, Springel2005} and defined as 
\begin{align}
  \frac{G m_J}{d_{iJ}^2} \pqty{\frac{b_J}{d_{iJ}}}^2 \leq \varDelta_\mathrm{acc} \abs{\vb*{a}_i^\mathrm{old}},
\end{align}
where $m_J$ and $b_J$ are the mass and the size of the group of distant particles, $d_{iJ}$ is the distance between the particle feeling the gravity and the distant particles causing the gravity, and $\vb*{a}_i^\mathrm{old}$ is the particle acceleration in the previous time step. 
This MAC enables a faster computation to achieve the same accuracy of the gravity calculation compared to other MACs as reported by \cite{Nelson2009, MikiUmemura2017}. 
\texttt{GOTHIC} generates a small interaction list shared by \num{32} concurrently working threads within a warp to achieve a high performance by increasing arithmetic intensity. 
When the MAC judges that the gravity calculation by a group of distant particles is acceptable, the corresponding pseudo particle is added to the interaction list. 
The MAC evaluation continues until the size of the list reaches its capacity, then calculates the gravitational force caused by the listed particles and flushes the list. 
This procedure continues until the calculation of gravity from all $N$-body particles finishes. 

The present study adjusts the gravitational octree code \texttt{GOTHIC} for the Volta GPU architecture, as well as measures and analyzes the performance of the code on Tesla V100 and Tesla P100. 
Section~\ref{sec:porting} summarizes the major revision in the Volta GPU architecture and introduces some recipes for the porting. 
Section~\ref{sec:results} presents the results of the performance measurements, and Section~\ref{sec:discussion} discusses the origin of the performance improvements in the Volta architecture compared to the Pascal architecture.

\section{Porting software to Volta GPU}
\label{sec:porting}
This section summarizes the major revision introduced in the Volta GPU architecture (\S\ref{subsec:cc70}) and the code tuning for Tesla V100 and Tesla P100 (\S\ref{subsec:microbench}). 

\subsection{Major revision in the Volta architecture}
\label{subsec:cc70}
The independent thread scheduling between \num{32} threads in a warp is the most significant revision in the Volta architecture compared to the earlier architectures. 
On GPUs with the Pascal or earlier architectures, all operations within \num{32} threads in a warp are executed at the same time; therefore, there is no need to execute explicit synchronization within a warp. 
On the other hand, simultaneity of operations is not guaranteed despite the operations are confined in \num{32} threads in a warp. 
Explicit synchronizations using \texttt{\_\_syncwarp()} instruction for \num{32} threads in a warp, or tiled synchronization using Cooperative Groups when the number of the involved threads is of the power of two and smaller than \num{32}, are necessary to proceed the operations properly. 

Removal of implicit synchronization within a warp introduces an adverse side effect in the warp divergence: it potentially extends the duration of the warp divergence. 
On GPUs based on earlier or equal to the Pascal architecture, the warp divergence occurs only within the conditional-branch sections, and all threads in a warp reconverge soon after the branch ends (Figure~20 of \cite{GV100Whitepaper}). 
On GPUs with the Volta architecture, the warp divergence lasts until executing any explicit synchronization even after the conditional-branch ended (Figures~22 and 23 of \cite{GV100Whitepaper}). 
Therefore, inserting explicit synchronization instructions at the appropriate location is necessary not to decelerate the computing on the Volta GPU. 
Since the procedure requires detailed reviewing and careful edit of the source code, NVIDIA provides a workaround without adding explicit synchronization. 
In particular, compilation with \texttt{-gencode arch=compute\_60,code=sm\_70} enforces the implicit synchronization of \num{32} threads within a warp similar to the Pascal architecture. 
Hereafter, we call the Pascal mode or the Volta mode for compilation with \texttt{-gencode arch=compute\_60,code=sm\_70} or \texttt{-gencode arch=compute\_70,code=sm\_70}, respectively. 
The Volta mode is the default mode in CUDA, while the Pascal mode alleviates the need of significant edit of the source code by taking a potential risk of dropping the degree of performance optimization by CUDA. 
In Sections~\ref{sec:results} and \ref{subsec:speedup_volta}, the Volta and Pascal modes are compared based on the time-to-solution. 

The absence of implicit synchronization within a warp also modifies the warp shuffle functions. 
NVIDIA deprecates functions like \texttt{\_\_shfl()} and \texttt{\_\_shfl\_xor()}, and encourages to use the warp-synchronous version of the functions like \texttt{\_\_shfl\_sync()} and \texttt{\_\_shfl\_xor\_sync()} instead. 
The new functions require a new input argument named mask to specify which thread should involve the function. 
For example, \texttt{0xffffffff} or \texttt{0xffff} are the appropriate mask when \num{32} threads or \num{16} threads in succession call the common warp shuffle function, respectively. 
Careful implementation is necessary if software execution has data dependency such as a gravitational octree code. 
Let us consider the case when \num{16} threads involve a warp shuffle function. 
There are two situations: (1) a single group of \num{16} threads calls the function or (2) two groups of \num{16} threads call the function at the same time. 
The expected mask for the former case is \texttt{0xffff} as usual; however, it is \texttt{0xffffffff} for the latter case. 
None of \texttt{0xffff} in both groups or \texttt{0xffff} for the first half-warp and \texttt{0xffff0000} for the second half-warp returns the correct value. 
Therefore, appropriately switching \texttt{0xffffffff} or \texttt{0xffff} depending on the case during the runtime is necessary to yield the correct results. 
Obtaining the appropriate mask during runtime is possible by calling \texttt{\_\_activemask()} just before calling the warp shuffle function. 
In the implementation of \texttt{GOTHIC} for the Volta GPU architecture, we set the mask as \texttt{0xffffffff} for the Pascal mode or the Volta mode when \num{32} threads never diverge, or the returned value by \texttt{\_\_activemask()}, when the \num{32} threads may diverge under the Volta mode. 

The Volta architecture introduces flexibility regarding the configuration of the on-chip memory compared to the earlier architectures. 
CUDA automatically determines the capacity of the shared memory per SM from the possible candidates of \num{0}, \num{8}, \num{16}, \num{32}, \num{64}, and \num{96}~\si{\kibi B}. 
Alternatively, user can manually specify the capacity by calling \texttt{cudaFuncSetAttribute()} with the name of the target function, \texttt{cudaFuncAttributePreferredSharedMemoryCarveout}, and the ratio of the expected capacity of the shared memory to the maximum value as integer. 
For example, inputting an integer value of \num{66} assigns \SI{64}{\kibi B}, which is $2/3 \sim 66.7$\% of \SI{96}{\kibi B}, as the shared memory. 
There is a pitfall that putting \num{67} assigns \SI{96}{\kibi B} instead of \SI{64}{\kibi B}; therefore, the input integer should be the largest integer value not greater than the expected ratio (i.e., the floor function of the value). 

CUDA 9 introduces global synchronization among multiple thread-blocks as a new feature by using Cooperative Groups. 
The function extends the flexibility of the implementation. 
In the case of \texttt{GOTHIC}, we use GPU lock-free synchronization proposed by \cite{XiaoFeng2010} instead of the global synchronization by Cooperative Groups because it has already been implemented and the micro-benchmarks show that the former version is faster than the latter one (see \ref{app:global.sync} for details).

\subsection{Tuning for Tesla V100 and Tesla P100}
\label{subsec:microbench}
\begin{table}[tb]
  \caption{Environments}
  \label{tab:env}
  \hbox to\hsize{\hfil
    \begin{tabular}{l|ll}\hline\hline
      CPU & IBM POWER9 (8335-GTG) & Intel Xeon E5-2680 v4\\
      & 16~cores, 2.0--\SI{3.1}{GHz} & 14~cores, \SI{2.4}{GHz}\\\hline
      GPU & Tesla V100 (SXM2) & Tesla P100 (SXM2)\\
       & 5120~cores, \SI{1.530}{GHz} & 3584~cores, \SI{1.480}{GHz}\\
      & HBM2 \SI{16}{GB} & HBM2 \SI{16}{GB}\\\hline
      Compiler & gcc 4.8.5 & icc 18.0.1.163\\
      & CUDA 9.2.88 & CUDA 8.0.61\\\hline
    \end{tabular}\hfil}
\end{table}
The performance of \texttt{GOTHIC} strongly depends on the configuration of the thread-block (e.g., the number of threads per thread-block \texttt{Ttot} for each function and the minimum number of threads involving common reduction operations or scan \texttt{Tsub}). 
Therefore, we performed micro-benchmarks to determine the optimal values of the fundamental parameters. 
The measurements were carried out on POWER9 servers at The University of Tokyo and TSUBAME3.0 supercomputer at Tokyo Institute of Technology. 
The POWER9 servers and TSUBAME3.0 are equipped with Tesla V100 and Tesla P100, respectively. 
Table~\ref{tab:env} lists the details of both environments. 

Since the performance of the tree code also depends on the particle distribution, we adopted a model of the Andromeda galaxy (M31) as a realistic distribution. 
The model is an updated version of the M31 model constructed by \cite{Geehan2006, Fardal2007}: a dark matter halo with the Navarro--Frenk--White model \cite{Navarro1995} (the mass is $\SI{8.11e11}{M_\odot}$ and the scale length is $\SI{7.63}{kpc}$), a stellar halo with the S\'ersic model \cite{Sersic1963} (the mass is $\SI{8e9}{M_\odot}$, the scale length is $\SI{9}{kpc}$, and the S\'ersic index is $2.2$ \cite{Gilbert2012, Ibata2014}), a stellar bulge with the Hernquist model \cite{Hernquist1990} (the mass is $\SI{3.24e10}{M_\odot}$ and the scale length is $\SI{0.61}{kpc}$), and an exponential disk (the mass is $\SI{3.66e10}{M_\odot}$, the scale length is $\SI{5.4}{kpc}$, the scale height is $\SI{0.6}{kpc}$, and the minimum of the Toomre's $Q$-value is $1.8$ \cite{Tenjes2017}). 
We utilized \texttt{MAGI} \cite{MikiUmemura2018} to generate the particle distribution in a dynamical equilibrium state under the constraint of the mass of all $N$-body particles being identical. 

\begin{table}[tb]
  \caption{Number of threads per thread-block}
  \label{tab:config}
  \hbox to\hsize{\hfil
    \begin{tabular}{l|rr|rr}\hline\hline
               & \multicolumn{2}{c|}{Tesla V100} & \multicolumn{2}{c}{Tesla P100}\\
      function & \texttt{Ttot} & \texttt{Tsub} & \texttt{Ttot} & \texttt{Tsub}\\\hline
      \texttt{walkTree} & 512 & 32 & 512 & 32\\
      \texttt{calcNode} & 128 & 32 & 256 & 16\\
      \texttt{makeTree} & 512 &  8 & 512 &  8\\
      \texttt{predict}  & 512 & -- & 512 & --\\
      \texttt{correct}  & 512 & 32 & 512 & 32\\\hline
    \end{tabular}\hfil}
\end{table}
We set the configuration of \texttt{GOTHIC} to achieve the shortest time-to-solution of \num{4096} steps integration using $N = 2^{23} = \num{8388608}$ particles with the accuracy controlling parameter $\varDelta_\mathrm{acc} = 2^{-9} = 1.953125 \times 10^{-3}$. 
Micro-benchmarks were carried out in the Volta mode on Tesla V100. 
Table~\ref{tab:config} summarizes the optimal configuration of \texttt{GOTHIC}: \texttt{walkTree} calculates the gravitational acceleration using the tree traversal, \texttt{calcNode} computes the location of the center-of-mass and the total mass of the tree nodes, \texttt{makeTree} builds the tree structure, \texttt{predict} and \texttt{correct} execute the orbit integration using the second-order Runge--Kutta method.

\section{Results}
\label{sec:results}
\begin{figure}[tb]
  \centering
  \includegraphics[width=\linewidth, pagebox=cropbox, clip]{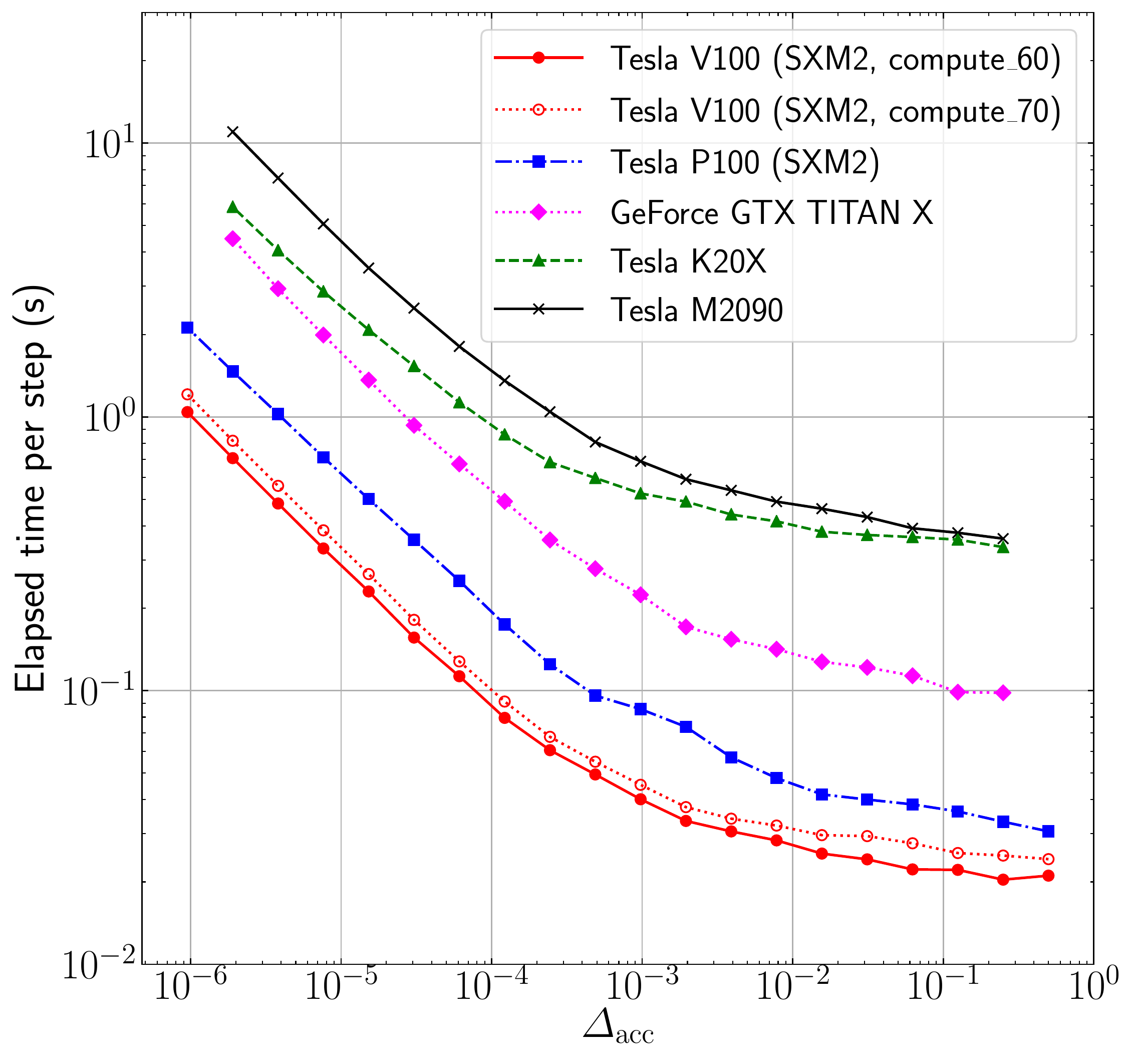}
  \caption{
    Execution time per step as a function of the accuracy controlling parameter $\varDelta_\mathrm{acc}$. 
    Each symbol indicates the measured results on different GPUs: red circles (Tesla V100), blue squares (Tesla P100), magenta triangles (GeForce GTX TITAN X), green triangles (Tesla K20X), and black crosses (Tesla M2090). 
    Red filled circles and red open circles exhibit results for the Pascal mode (compute\_60) and the Volta mode (compute\_70), respectively. 
    Measurements on GPUs with the Maxwell architecture or older GPUs (GeForce GTX TITAN X, Tesla K20X, and Tesla M2090) were carried out by \cite{MikiUmemura2017}. 
  }
  \label{fig:m31_block_time}
\end{figure}
We measured the execution time as a function of the accuracy controlling parameter $\varDelta_\mathrm{acc}$ from $2^{-1}$ to $2^{-20}$ on Tesla P100 and Tesla V100. 
We compared the results with the Volta mode and the Pascal mode on Tesla V100. 
The other settings of the performance measurements are identical to the micro-benchmarks in \S\ref{subsec:microbench}. 
The measured execution time with the typical accuracy of $\varDelta_\mathrm{acc} = 2^{-9} = 1.953125 \times 10^{-3}$ are \SIlist[list-final-separator = {, and }]{7.4e-2;3.8e-2;3.3e-2}{s} on Tesla P100, Tesla V100 in the Volta mode, and Tesla V100 in the Pascal mode, respectively. 
Figure~\ref{fig:m31_block_time} shows the results of the measurements. 
The figure also compiles the results measured previously \cite{MikiUmemura2017} on earlier GPUs (GeForce GTX TITAN X, Tesla K20X, and Tesla M2090). 
The execution time on later GPUs is always faster than that of earlier GPUs. 
The observed dependence of the execution time on the accuracy controlling parameter is similar in all architectures except for the Kepler architecture. 
Figure~\ref{fig:m31_block_time} confirms that Tesla V100 achieves an acceleration that is tenfold higher than that of Tesla M2090 in the same algorithm. 
On Tesla V100, the Pascal mode is always faster than the Volta mode. 
Hereafter, we regard the Pascal mode as a fiducial configuration on Tesla V100. 

\begin{figure}[tb]
  \centering
  \includegraphics[width=\linewidth, pagebox=cropbox, clip]{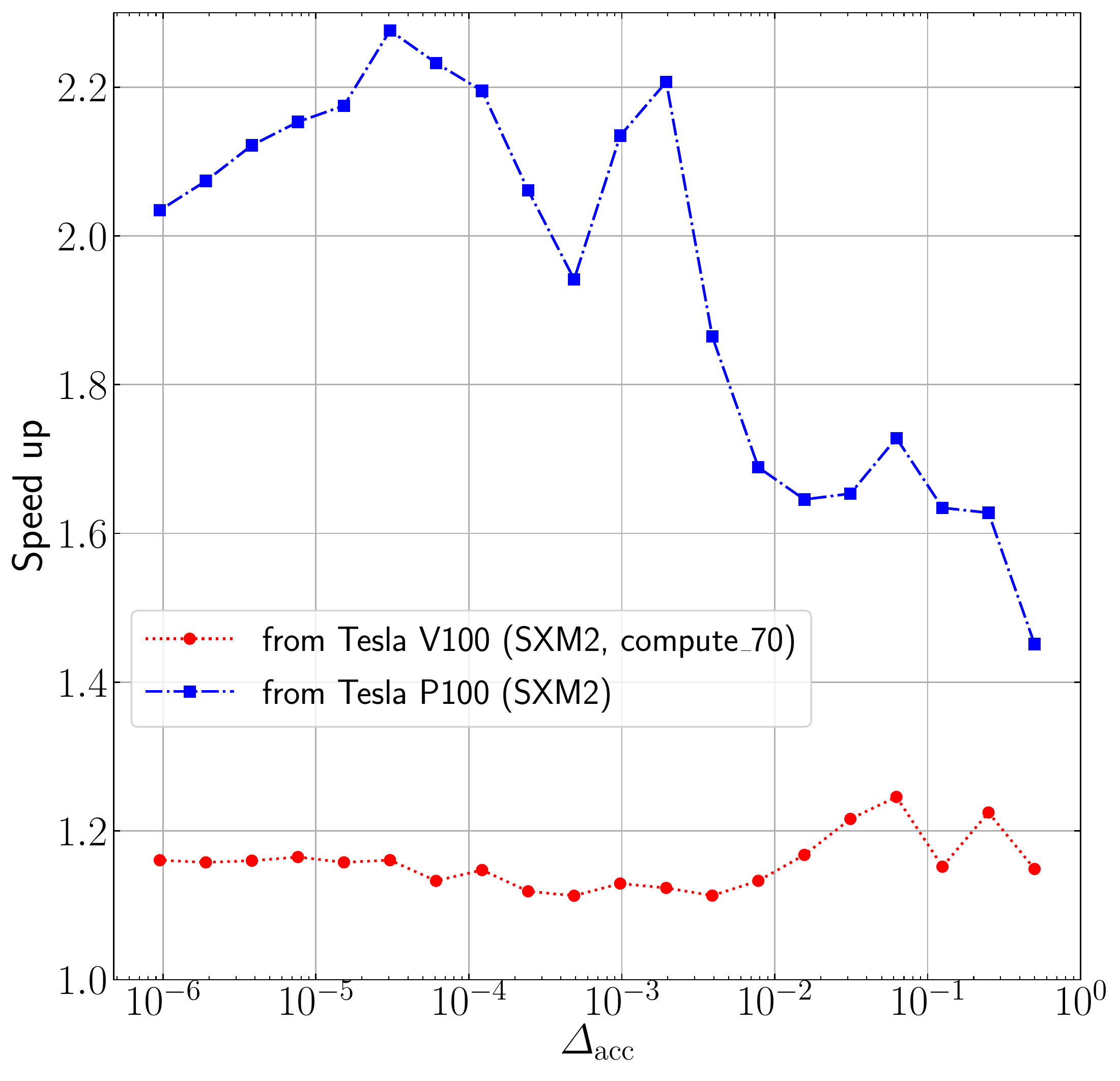}
  \caption{
    Speed-up of \texttt{GOTHIC} as a function of the accuracy controlling parameter $\varDelta_\mathrm{acc}$. 
    Red circles and blue squares show the speed-up of Tesla V100 with the Pascal mode compared to Tesla V100 with the Volta mode and Tesla P100, respectively. 
  }
  \label{fig:m31_block_gen}
\end{figure}
Figure~\ref{fig:m31_block_gen} shows the speed-up of Tesla V100 in the Pascal mode compared to Tesla P100 or Tesla V100 in the Volta mode. 
The Pascal mode is \numrange[range-phrase = --]{1.1}{1.2} times faster than the Volta mode irrespective of the accuracy of the gravity calculation (the red circles with the dotted curve). 
Tesla V100 achieves a \numrange{1.4}{2.2}-fold acceleration compared to Tesla P100 (the blue squares with the dot-dashed curve). 
In particular, the speed-up ratio exceeds \num{2} in the regions of $\varDelta_\mathrm{acc} \lesssim 10^{-3}$. 
The observed value is greater than \num{1.5}, which is the ratio of the theoretical peak performance of Tesla V100 to that of Tesla P100. 
The origin of the higher speed-up than the GPU performance improvements is discussed in \S\ref{subsec:speedup_pascal}. 

\begin{figure}[tb]
  \centering
  \includegraphics[width=\linewidth, pagebox=cropbox, clip]{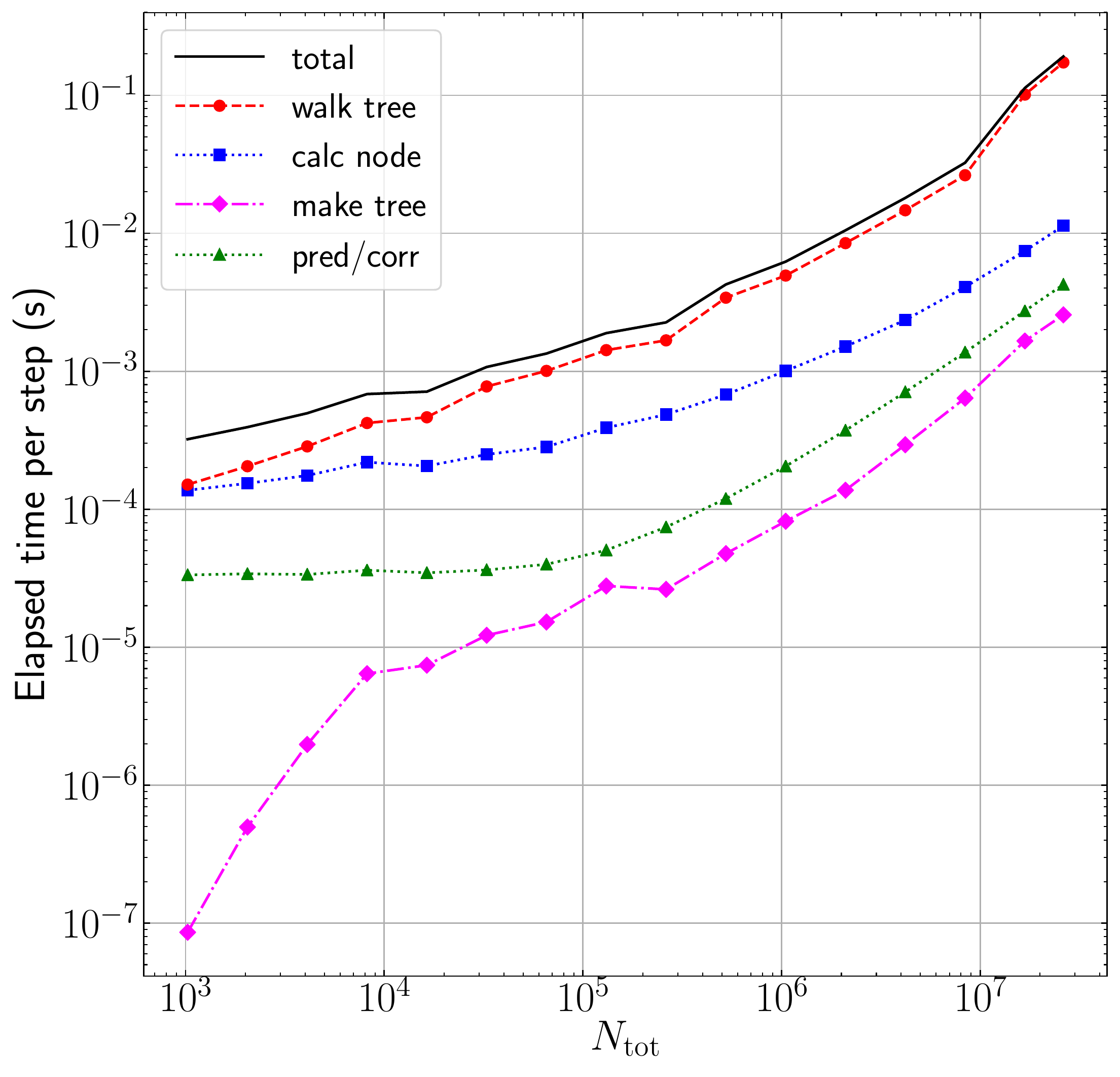}
  \caption{
    Dependence of the elapsed time of \texttt{GOTHIC} on the total number of $N$-body particles $N_\mathrm{tot}$. 
    Each symbol shows the elapsed time of the representative functions: red circles with dashed curve (gravity calculation), blue squares with dotted curve (calculating the location of the center-of-mass and the total mass of the tree nodes), magenta diamonds with dot-dashed curve (tree construction), green triangles with dotted curve (orbit integration), and black solid curve (sum of them). 
  }
  \label{fig:m31_ndep}
\end{figure}
We also measured the execution time of \texttt{GOTHIC} with varying the total number of $N$-body particles $N_\mathrm{tot}$ from $2^{10} = \num{1024}$ to $25 \times 2^{20} = \num{26214400}$ on Tesla V100. 
The number of time steps in the measurements is \num{65536} to ensure the accuracy of the measurements using small particle numbers. 
The execution time per step with the maximum number of particles is \SI{2.0e-1}{s}. 
Figure~\ref{fig:m31_ndep} exhibits the scaling of \texttt{GOTHIC} to $N_\mathrm{tot}$ and the breakdown of the representative functions. 
Similar to all earlier GPU architectures, the execution time for gravity calculation (the red circles with the dashed curve) is always the dominant contributor; however, the contribution from \texttt{calcNode} (the blue squares with the dotted curve) is not negligible especially in the small $N_\mathrm{tot}$ regions. 

On Tesla P100, the identical measurements reveal that the execution time is \SI{3.3e-1}{s} for the maximum particle number of $30 \times 2^{20} = \num{31457280}$. 
The reason for Tesla P100 can perform larger simulations than Tesla V100 is the higher capacity of the global memory per SM. 
Both GPUs have \SI{16}{GB} HBM2 memory, but Tesla V100 has $\sim 1.4$ times more SMs. 
Since \texttt{GOTHIC} adopts the breadth-first traverse and every SM needs the buffer to store tree cells to be evaluated, the buffer size per SM determines the maximum problem size. 
Therefore, Tesla V100 with \SI{32}{GB} HBM2 would be able to perform $N$-body simulations using more particles than Tesla P100.

\section{Discussion}
\label{sec:discussion}
\subsection{Origin of speed-up from the Volta mode}
\label{subsec:speedup_volta}
Performance measurements presented in the previous section reveal that the Pascal mode is \numrange{1.1}{1.2} times faster than the Volta mode. 
The Pascal mode disables the independent thread scheduling of the Volta GPU architecture and certifies that \num{32} threads in a warp execute operations simultaneously, similar to the Pascal or earlier GPU architectures. 
It reduces overheads by \texttt{\_\_syncwarp()} instruction or tiled synchronization by Cooperative Groups. 
On the other hand, it disables some features introduced in the Volta architecture and also has a risk to drop the performance due to insufficient performance optimization for the Volta GPU architecture by the CUDA compiler. 

\begin{figure}[tb]
  \centering
  \includegraphics[width=\linewidth, pagebox=cropbox, clip]{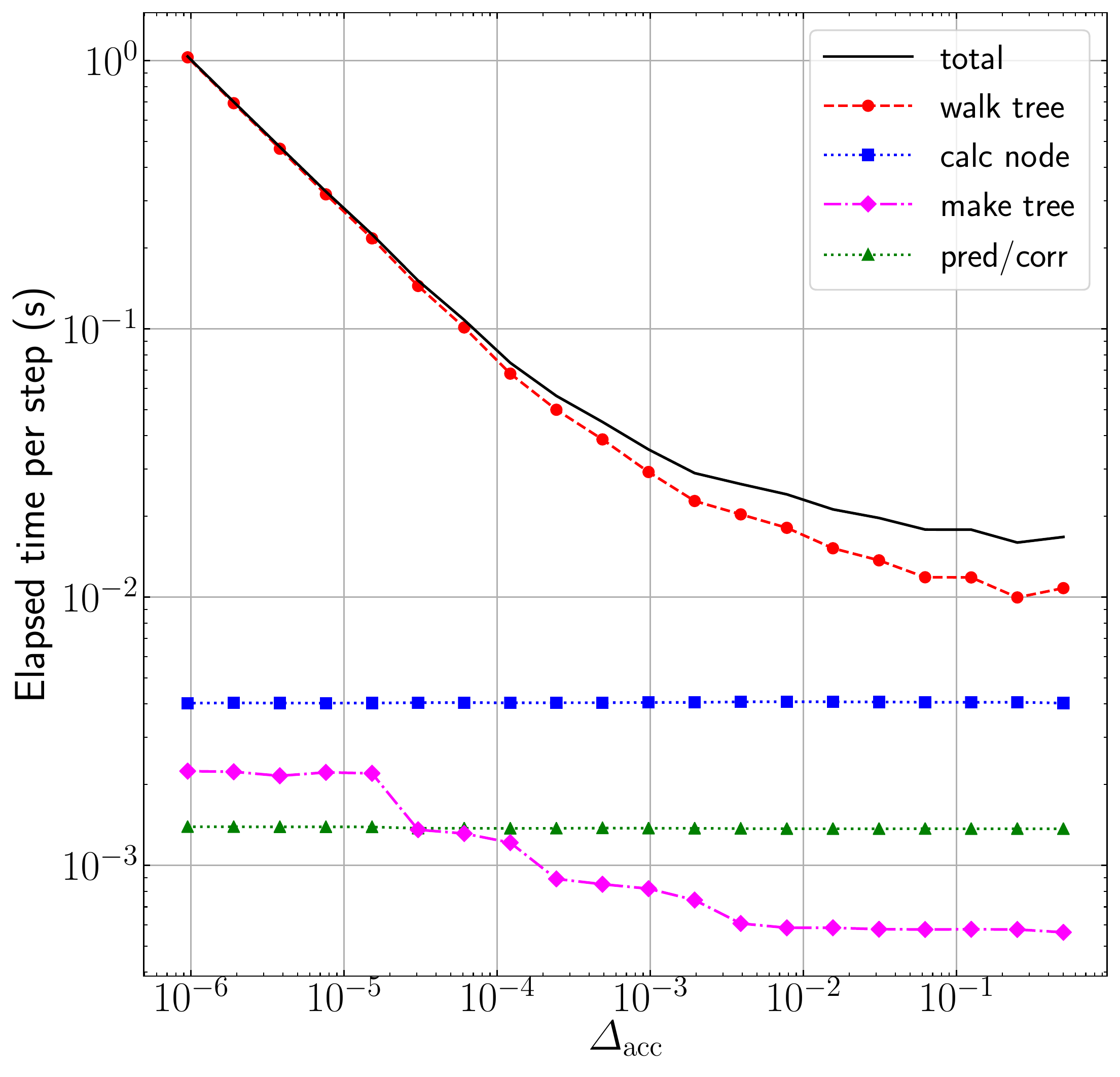}
  \caption{
    Breakdown of the elapsed time of \texttt{GOTHIC} as a function of the accuracy controlling parameter $\varDelta_\mathrm{acc}$. 
    Symbols and lines are the same as those in Figure~\ref{fig:m31_ndep}. 
  }
  \label{fig:m31_frac}
\end{figure}
To evaluate the pros and cons of using the Pascal mode, we measured the execution time of the representative functions in \texttt{GOTHIC} and compared them between the Pascal and the Volta modes. 
Figure~\ref{fig:m31_frac} presents the execution time of the representative functions in \texttt{GOTHIC} on Tesla V100 with the Pascal mode at $N = 2^{23} = \num{8388608}$. 
The figure reveals that the execution time of \texttt{walkTree} decreases with the decrease of the accuracy of gravity calculation (the red circles with the dashed curve), while that of \texttt{calcNode} (the blue squares with the dotted line) or orbit integration (the green triangles with the dotted line) are independent of the accuracy of gravity calculation. 
It is a natural consequence, because the precise computation of gravity takes a longer time compared to the low-accurate computation, while computation of tree nodes and orbit integration are irrelevant for the accuracy of the gravitational force. 

The tree construction is essentially independent from the accuracy of gravity calculation; however, the interval of tree rebuilds reflects the change of the execution time for gravity calculation since \texttt{GOTHIC} automatically adjusts the tree rebuild interval to minimize the sum of the execution time for gravity calculation and tree construction. 
Throughout the measurements, the rebuild interval is six steps in the highest accuracy and around \num{30} steps in the lowest accuracy. 

\begin{figure}[tb]
  \centering
  \includegraphics[width=\linewidth, pagebox=cropbox, clip]{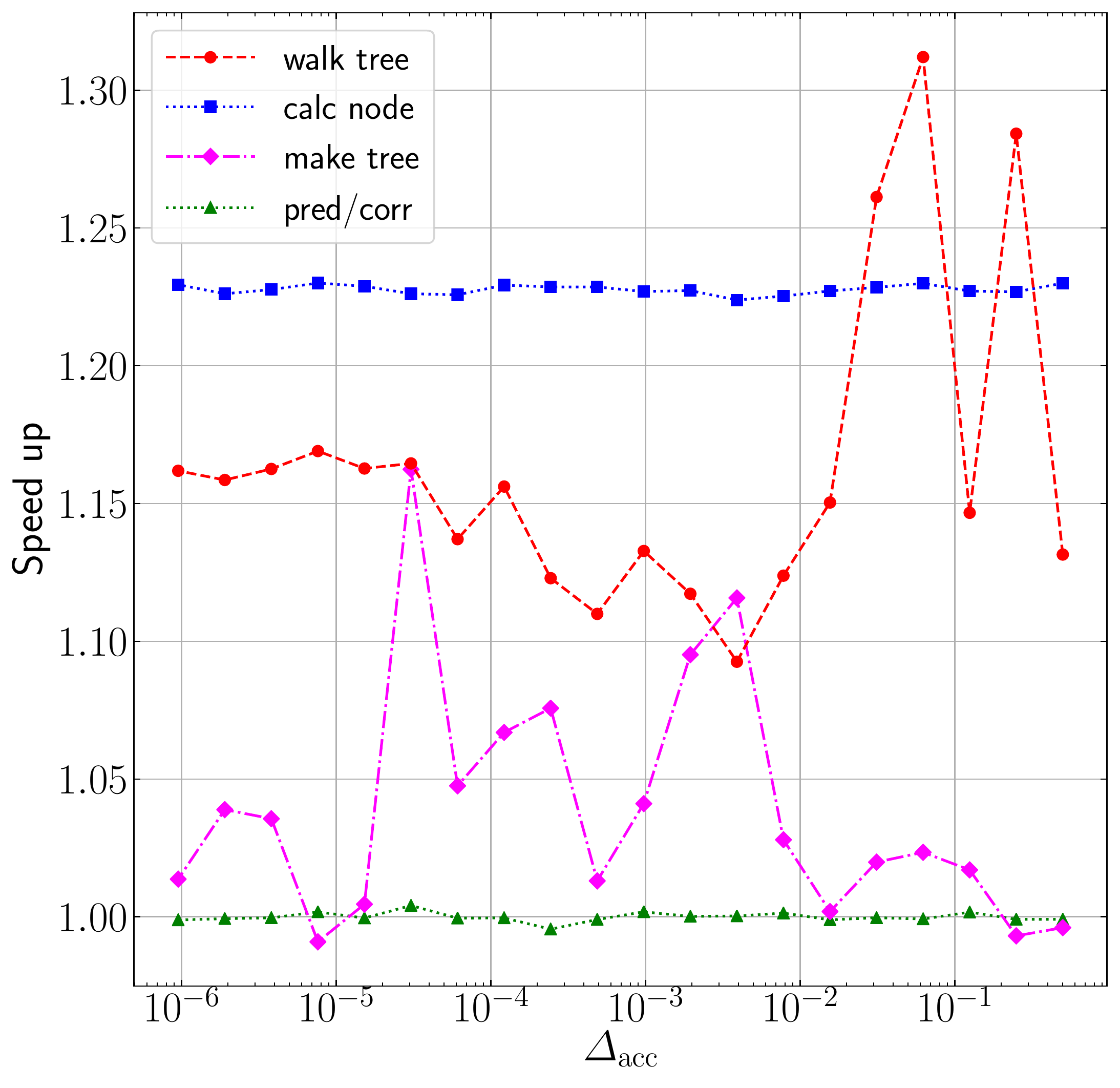}
  \caption{
    Speed-up of the Pascal mode compared to the Volta mode as a function of the accuracy controlling parameter $\varDelta_\mathrm{acc}$. 
    Symbols and lines are the same as those in Figure~\ref{fig:m31_ndep}. 
  }
  \label{fig:m31_mode}
\end{figure}
Figure~\ref{fig:m31_mode} shows the speed-up ratio of the Pascal mode compared to the Volta mode for the representative functions in \texttt{GOTHIC}. 
The figure indicates that the Pascal mode provides faster calculation for all functions. 
In case of the dominant contributor, \texttt{walkTree}, the Pascal mode is about \num{15}\% faster and tends to increase the speed-up rate in $\varDelta_\mathrm{acc} \gtrsim 10^{-2}$. 
Furthermore, \texttt{calcNode}, which shows about \num{23}\% acceleration, contributes to the overall speed-up in the regions of $\varDelta_\mathrm{acc} \gtrsim 10^{-2}$. 

The above two functions are the origin of the observed \numrange{1.1}{1.2}-fold acceleration of the Pascal mode compared to the Volta mode. 
Both functions include reduction operations and scan within \num{32} threads consisting a warp, and therefore, call \texttt{\_\_syncwarp()} for many times in the Volta mode. 
Since the Pascal mode never calls \texttt{\_\_syncwarp()}, the origin of the speed-up is considered as the execution cost of \texttt{\_\_syncwarp()}. 
It implies that the Pascal mode will achieve better performance for functions having many synchronizations within a warp. 

Orbit integration does not require inner-warp synchronization using \texttt{\_\_syncwarp()}, and operations would be identical in both the Pascal and the Volta modes. 
The only possible difference between the two modes is the degree of performance optimization by the CUDA compiler. 
As the figure shows, there is no difference in the performance between the two modes, which implies that the degree of the optimization is similar in the two modes. 

The function for tree construction, which executes tiled synchronization by Cooperative Groups and \texttt{\_\_activemask()}, does not show a higher speed-up ratio compared to the functions using \texttt{\_\_syncwarp()} only. 
The probable reasons are that the function also exploits intra-block synchronization and most of the execution time is the radix sort using \texttt{cub::DeviceRadixSort::SortPairs} function in CUB \cite{CUB}. 
Both reduce the contribution of the synchronization costs within a warp.

\subsection{Origin of speed-up from Tesla P100}
\label{subsec:speedup_pascal}
This section elucidates what enables the observed speed-up to exceed the ratio of the theoretical peak performance between Tesla V100 and Tesla P100. 
We concentrate on the function for gravity calculation since it is the dominant contributor to the execution time. 
The independence of the integer operation units from the floating-point number operation units on GPUs with the Volta architecture introduces opportunities to hide the execution time of integer operations or floating-point number operations by implementing their overlapped execution. 
The acceleration due to the overlapping is only available when both integer and floating-point number operations are executed independently in compute-intensive situations. 
$N$-body simulations are the representative examples of compute-intensive applications, where the tree method includes integer operations, as opposed to the direct method, which executes floating-point number operations only. 
The gravitational octree simulations are the ideal testbed to confirm whether the acceleration by hiding integer operations is available or not in the actual applications. 

To examine the hypothesis that the overlapping of integer and floating-point number operations achieves the observed speed-up, we counted the number of instructions executed in \texttt{walkTree} using \texttt{nvprof}. 
The metrics specified in the measurements are \texttt{inst\_integer}, \texttt{flop\_count\_sp\_fma}, \texttt{flop\_count\_sp\_add}, \texttt{flop\_count\_sp\_mul}, and \texttt{flop\_count\_sp\_special}. 
The measurements using \texttt{nvprof} with the above metrics serialize the execution of \texttt{GOTHIC} and the computations become significantly slow. 
Hence, we ran the simulation with $N = 2^{23} = \num{8388608}$, $\varDelta_\mathrm{acc} = 2^{-9} = 1.953125 \times 10^{-3}$ and \num{256} steps on Tesla V100. 
We also disabled the auto-tuning of the tree reconstruction intervals and set a fixed value using the interval taken from the executions without \texttt{nvprof} considering that \texttt{nvprof} extends the execution time of the functions that are the basis of the auto-tuning. 

\begin{figure}[tb]
  \centering
  \includegraphics[width=\linewidth, pagebox=cropbox, clip]{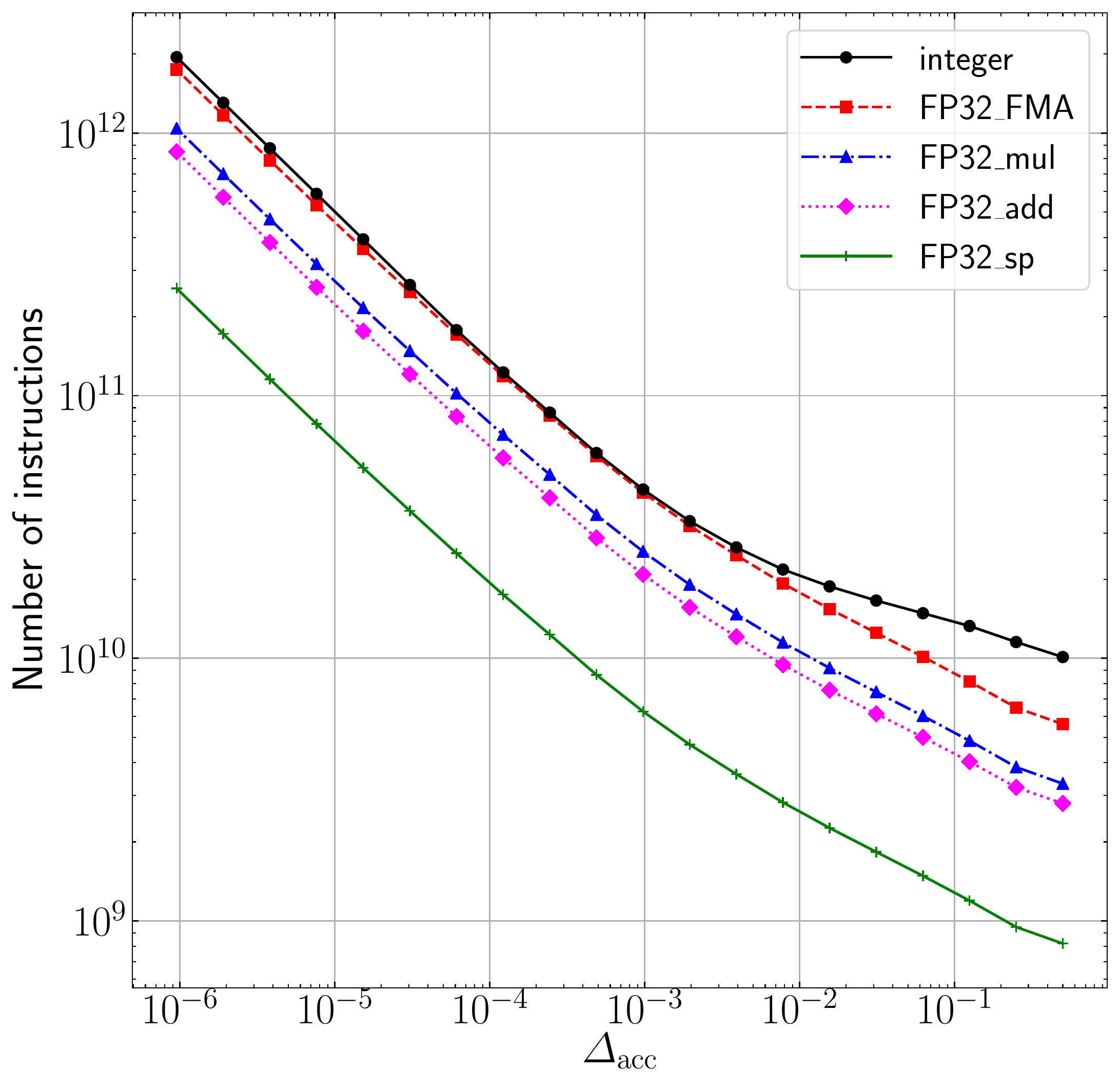}
  \caption{
    Number of instructions per step executed in the function for gravity calculation. 
    Every symbol with curves reveals the measured results with different metrics: black circles with solid curve (integer), red squares with dashed curve (single-precision FMA operations), blue triangles with dot-dashed curve (single-precision multiplications), magenta diamonds with dotted curve (single-precision additions), and green pluses with solid curve (single-precision reciprocal square root). 
  }
  \label{fig:m31_inst}
\end{figure}
Figure~\ref{fig:m31_inst} shows the measured number of instructions as a function of the accuracy controlling parameter. 
The number of instructions executed by the special function units (the green crosses with the solid curve), which are the reciprocal square root of single-precision floating-point numbers, is nearly tenfold smaller than that of the fused multiply-add (FMA) operations (the red squares with the dashed curve). 
Even considering the throughput of the reciprocal square root to be a quarter of the throughput of the operations executed by CUDA cores, the execution time of the reciprocal square root is smaller than that of the other floating-point number operations. 
For simplicity, we hereafter assume that the execution time of the reciprocal square root is entirely hidden by the other floating-point number operations. 

\begin{figure}[tb]
  \centering
  \includegraphics[width=\linewidth, pagebox=cropbox, clip]{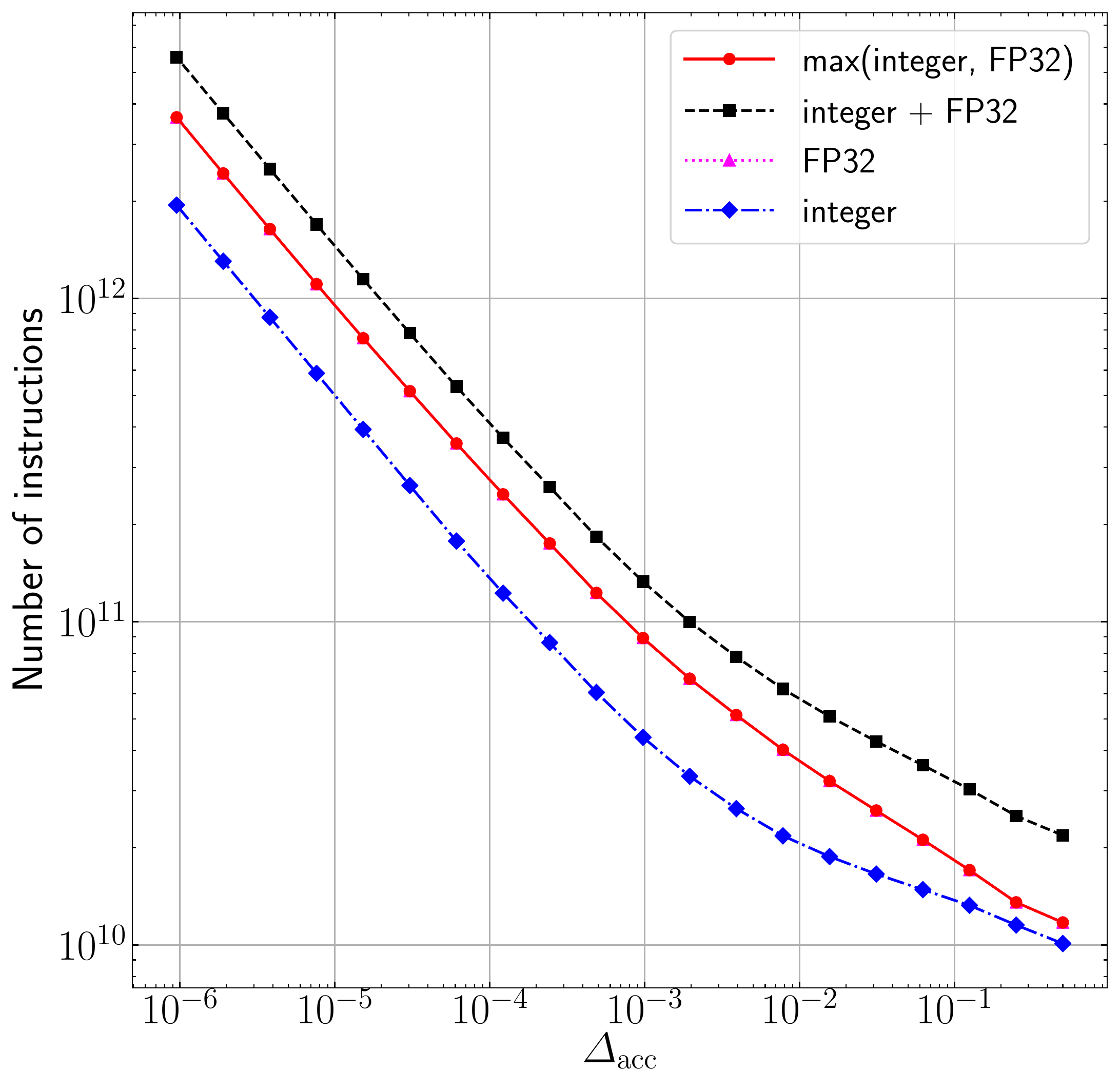}
  \caption{
    Number of instructions for different operating units. 
    The number of single-precision floating-point number instructions (magenta triangles with dotted curve, sum of FP32\_FMA, FP32\_mul, and FP32\_add in Figure~\ref{fig:m31_inst}) is always greater than that of integer instructions (blue triangles with dot-dashed curve, integer in Figure~\ref{fig:m31_inst}) and is identical to the maximum counts of the floating-point number and integer instructions (red circles with solid curve). 
    Black squares with dashed curve indicate the summed counts of the floating-point number and integer instructions. 
  }
  \label{fig:m31_op}
\end{figure}
Figure~\ref{fig:m31_op} compares the number of instructions executed on different operating units. 
The operation count of integer operations (the blue squares with the dot-dashed curve) is identical to that of Figure~\ref{fig:m31_inst}. 
The sum of all floating-point number operations performed by CUDA cores (FMA, multiplication, and addition) is plotted using the magenta triangles with the dotted curve. 
The instruction counts of floating-point number operations are always higher than that of integer operations; hence, the maximum number of both instructions (the red circles with the solid curve) corresponds to that of floating-point number operations. 
It implies that the execution time of integer operations can be hidden by the overlapped execution of the floating-point number operations on GPUs with the Volta architecture if there are no obstacles such as the dependency between the two types of operations. 
The black squares with the dashed curve indicate the summed number of integer and floating-point number operations. 
The summed number determines the execution time when the identical operation unit executes integer operations and floating-point number operations similar to GPUs with the Pascal or earlier architectures. 

\begin{figure}[tb]
  \centering
  \includegraphics[width=\linewidth, pagebox=cropbox, clip]{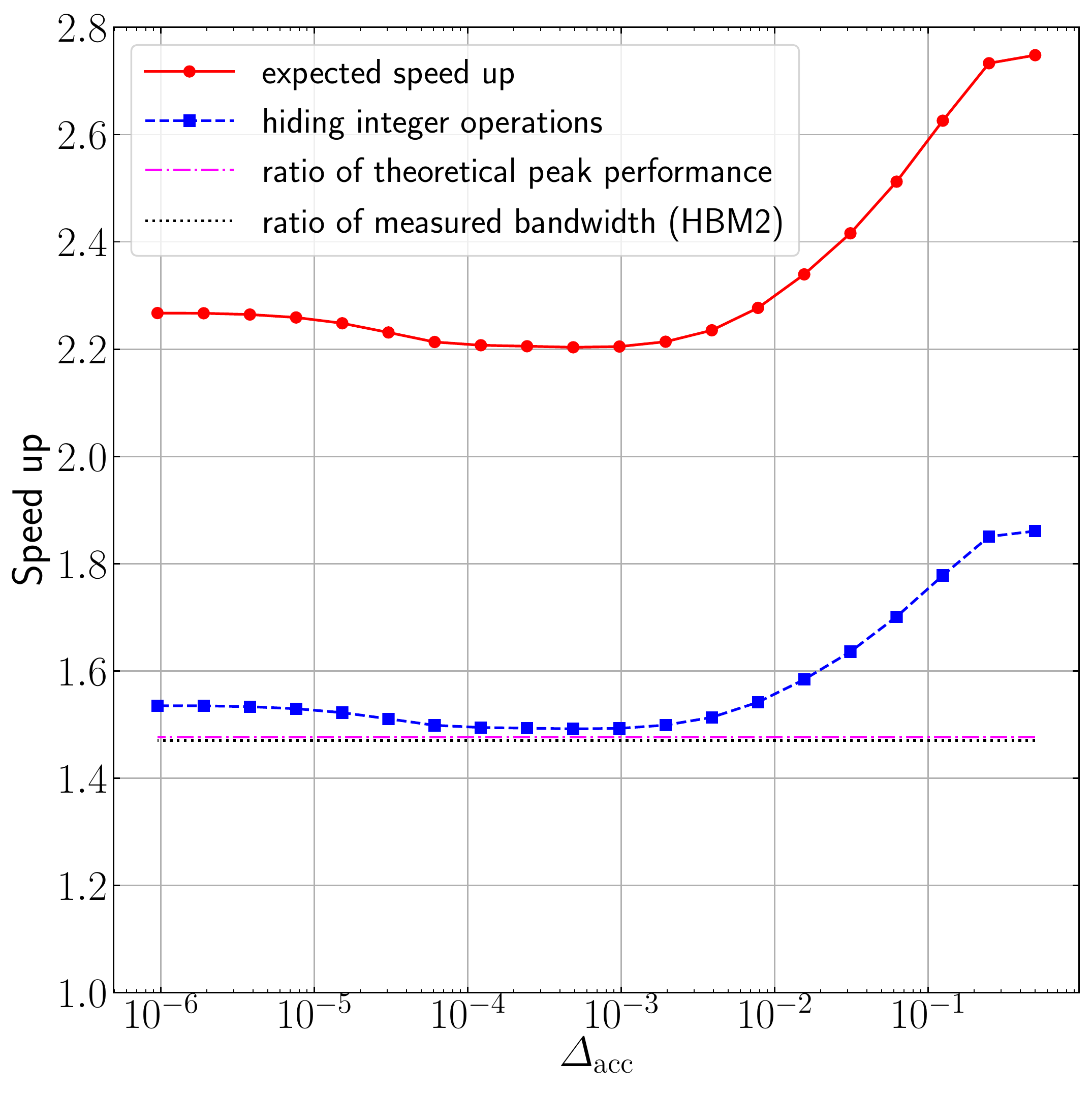}
  \caption{
    Expected speed-up of Tesla V100 compared to Tesla P100. 
    Magenta dot-dashed line and black dotted line indicate the ratio of the theoretical peak performance and the measured memory bandwidth of Tesla V100 compared to Tesla P100, respectively. 
    Blue squares with dashed curve show the expected speed-up by hiding integer operations (the ratio of black squares to red circles in Figure~\ref{fig:m31_op}). 
    The multiplication of blue squares with dashed curve and magenta line is the theoretical expectation of the observed speed-up (red circles with solid curve). 
  }
  \label{fig:m31_speedup}
\end{figure}
Figure~\ref{fig:m31_speedup} is the theoretical prediction of the speed-up ratio of Tesla V100 compared to Tesla P100. 
The magenta dot-dashed and the black dotted lines represent the ratio of the theoretical peak performance and the measured memory bandwidth, respectively. 
The blue squares with the dashed curve represent the ratio of the instructions, in which the execution of integer operations is never hidden (corresponding to Tesla P100, the black squares in Figure~\ref{fig:m31_op}) to the instructions, in which the execution of integer operations is completely hidden (the ideal situation on Tesla V100, the red circles in Figure~\ref{fig:m31_op}). 
The actual prediction is the product of the theoretical peak performance ratio and the ratio of the two scenarios (the red circles with the solid curve). 
The theoretical prediction suggests that the observed \num{2.2} fold acceleration is possible in $\varDelta_\mathrm{acc} \lesssim 10^{-3}$. 
On the other hand, the model fails to explain the observed decline of the speed-up in $\varDelta_\mathrm{acc} \gtrsim 10^{-3}$. 

Possible explanations for the disagreement between the theoretical prediction and the observed speed-up in $\varDelta_\mathrm{acc} \gtrsim 10^{-3}$ include the following: (1) there is no room for the overlapped execution of integer and floating-point number operations; (2) the transition of the performance limiting factor from compute-bound to memory-bound. 
The currently focused regions are the low-accuracy regions for gravity calculation; therefore, MAC evaluations tend to judge $N$-body particles causing the gravity to be sufficiently far and calculate the gravitational force using the tree nodes in the levels nearby the root node. 
To enable the overlapped execution of the integer and floating-point number operations, multiple iterations of the procedure that fills the interaction list in the shared memory by MAC evaluations (dominated by integer operations) and flushes the lists just after the gravity calculations (dominated by floating-point number operations) are necessary. 
Decreasing the accuracy of the gravity calculation might reduce the chance of the overlapped execution of the two types of operations, and therefore, the observed speed-up ratio goes down compared to the ratio of the theoretical peak performance. 
Reduction of opportunities for the overlapping also makes it difficult to hide the access latency in the global memory. 
It leads the compute-bound kernel to become the latency-bound kernel, which means that the overlapping of multiple executions cannot contribute to the total execution time. 
Furthermore, the decrease of the executed operations may increase the byte-per-flop ratio, leading to the compute-bound kernel to be converted to the bandwidth-bound function. 

\begin{figure}[tb]
  \centering
  \includegraphics[width=\linewidth, pagebox=cropbox, clip]{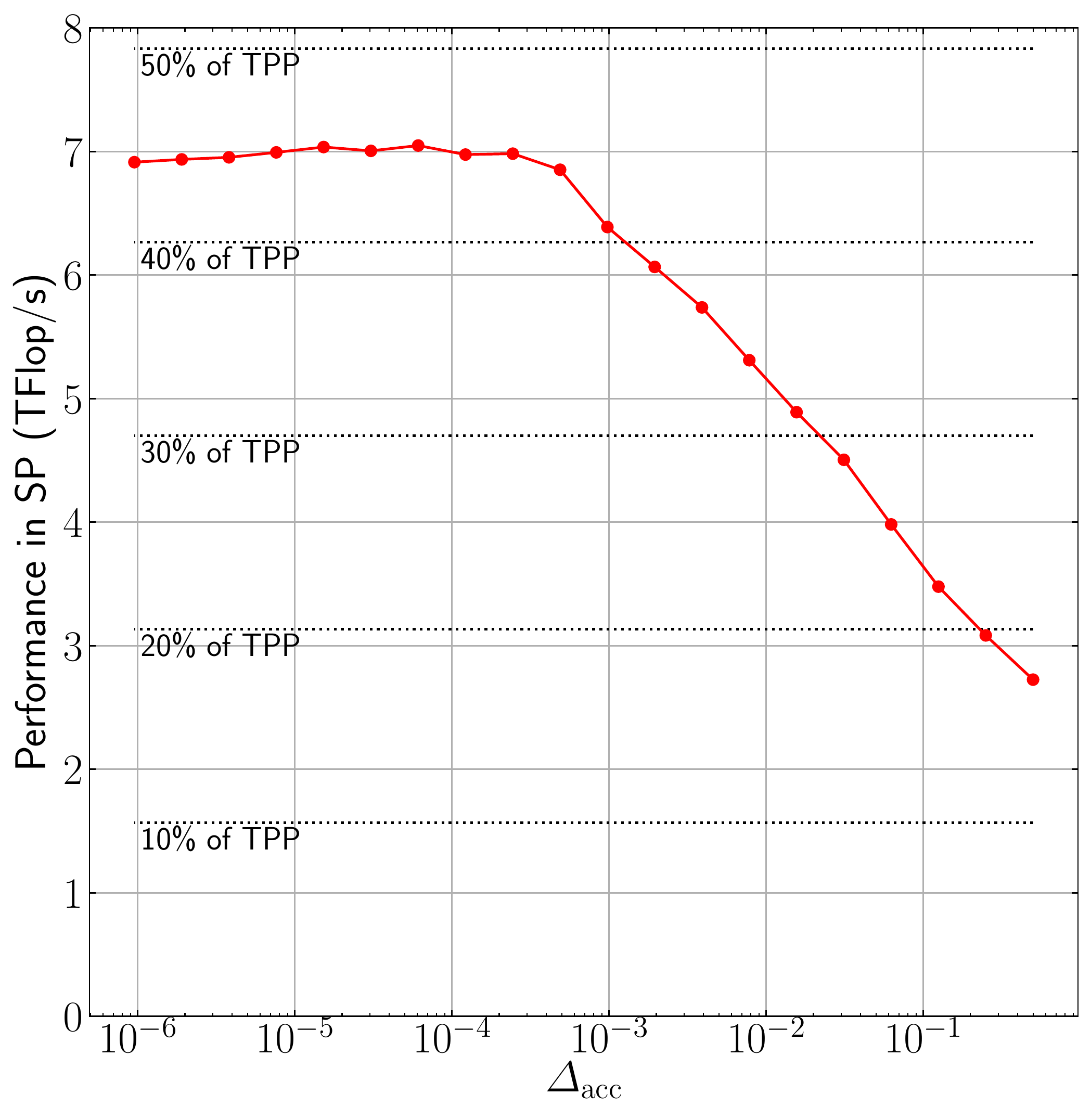}
  \caption{Measured performance of the function for gravity calculation as a function of the accuracy controlling parameter $\varDelta_\mathrm{acc}$. }
  \label{fig:m31_flops_walk}
\end{figure}
The direct inspection of the reason for the disagreement between the theoretical model and the observed speed-up is difficult. 
However, a performance drop would be observed when the force accuracy is decreased providing that some of the above considerations are correct. 
The performance evaluation requires individual measurements of the execution time of the gravity calculation function and the assumption of the \si{Flops} count of the single-precision reciprocal square root. 
Here we assume that the reciprocal square root corresponds to four \si{Flops} since the ratio of the throughput of this operation to the single-precision addition or multiplication operation is four \cite{Miki2013, MikiUmemura2017}. 
Figure~\ref{fig:m31_flops_walk} reveals the estimated performance of the gravity calculation. 
The sustained performance reaches \SI{7}{TFlop/s}, which is \num{45}\% of the single-precision theoretical peak performance, in the regions of $\varDelta_\mathrm{acc} \lesssim 10^{-3}$. 
The performance decreases with the increase of the accuracy controlling parameter as shown in Figure~\ref{fig:m31_flops_walk}. 
This suggests that the reduced computational workload with the decrease of the force accuracy deteriorate the performance and the speed-up ratio. 

\begin{figure}[tb]
  \centering
  \includegraphics[width=\linewidth, pagebox=cropbox, clip]{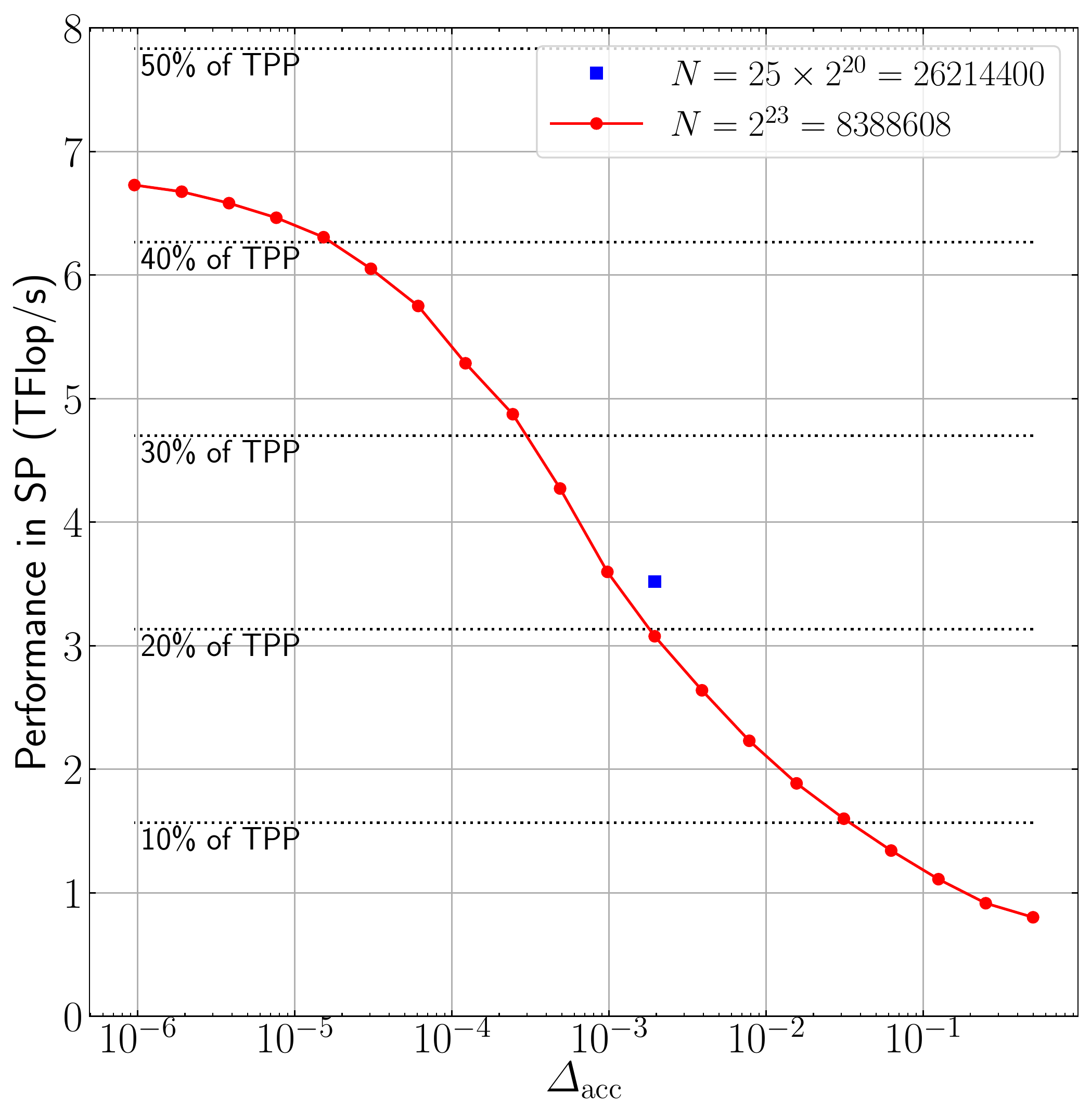}
  \caption{
    Measured performance of \texttt{GOTHIC} as a function of the accuracy controlling parameter $\varDelta_\mathrm{acc}$. 
    Red circles with solid curve and blue square represent the measured performance, where the total number of $N$-body particles is $N = 2^{23} = \num{8388608}$ and $N = 25 \times 2^{20} = \num{26214400}$, respectively. 
  }
  \label{fig:m31_flops_tot}
\end{figure}
The division of the obtained floating-point number operations by the total elapsed time of the code indicates the estimated performance of \texttt{GOTHIC} (Figure~\ref{fig:m31_flops_tot}). 
The measured single-precision performance is \SI{3.1}{TFlop/s} (\num{20}\% of the theoretical peak performance) and \SI{3.5}{TFlop/s} (\num{22}\% of the theoretical peak performance) for $N = 2^{23} = \num{8388608}$ and $N = 25 \times 2^{20} = \num{26214400}$, respectively, at $\varDelta_\mathrm{acc} = 2^{-9} = 1.953125 \times 10^{-3}$. 
The performance dependency on $\varDelta_\mathrm{acc}$ becomes stronger than that of in Figure~\ref{fig:m31_flops_walk} because the execution time of \texttt{calcNode} also contributes to the total execution time in the regions of greater $\varDelta_\mathrm{acc}$.

\section{Summary}
\label{sec:summary}
The gravitational octree code \texttt{GOTHIC}, which was originally optimized for GPUs with the Fermi, Kepler, and Maxwell architectures, was adapted to the Volta architecture. 
The Volta architecture introduces the independent thread scheduling and removes implicit synchronization within \num{32} threads consisting a warp. 
The change of the programming model requires modifying the code by inserting explicit synchronizations in appropriate locations or enforcing execution with implicit synchronization similar to the Pascal or earlier architectures by specifying \texttt{-gencode arch=compute\_60,code=sm\_70} during the compilation. 

The performance on Tesla V100 and Tesla P100, which are the latest and the previous flagship GPUs developed by NVIDIA, respectively, was measured. 
The performance measurements on Tesla V100 revealed that the $N$-body simulations of the Andromeda galaxy model with $2^{23} = \num{8388608}$ particles take \SI{3.8e-2}{s} with explicit synchronizations or \SI{3.3e-2}{s} in the same execution mode as that of the Pascal or earlier architectures. 
The results showed that the latter case is faster regardless of the gravity calculation accuracy, while specifying \texttt{-gencode arch=compute\_60,code=sm\_70} enables a \numrange{1.1}{1.2}-fold acceleration. 

Tesla V100 achieves a \numrange{1.4}{2.2}-fold acceleration compared to Tesla P100. 
The observed speed-up of \num{2.2} is higher than the improvement of the theoretical peak performance. 
The independence of the integer operation units from the floating-point number operation units enables the overlapped execution of integer operations and floating-point number operations. 
Such overlapped execution hides the execution time of the integer operations and leads to the speed-up rate above the theoretical peak performance ratio. 

Tesla V100 can execute the $N$-body simulation with up to $25 \times 2^{20} = \num{26214400}$ particles and takes \SI{2.0e-1}{s} per step. 
It corresponds to \SI{3.5}{TFlop/s} which is \num{22}\% of the single-precision theoretical peak performance.

\section*{Acknowledgment}
This work is supported by ``Joint Usage/Research Center for Interdisciplinary Large-scale Information Infrastructures'' and ``High Performance Computing Infrastructure'' in Japan (Project ID: jh180045-NAH), and ``TSUBAME Encouragement Program for Young/Female Users'' of Global Scientific Information and Computing Center, Tokyo Institute of Technology. 
The performance optimization for Tesla P100 and performance measurements using Tesla P100 were carried out on the TSUBAME3.0 supercomputer at Tokyo Institute of Technology.

\appendix

\section{Global synchronization using Cooperative Groups}
\label{app:global.sync}
CUDA 9 introduces a global synchronization function among multiple thread-blocks using the Cooperative Groups as a new feature. 
\texttt{GOTHIC} is already equipped with global synchronization using GPU lock-free synchronization proposed by \cite{XiaoFeng2010} since the code had been developed before the Cooperative Groups was introduced. 

To exploit global synchronization using the Cooperative Groups when the source code is split into multiple files, it is necessary to append \texttt{--device-c} and \texttt{--device-link --output-file tmp.o} while generating object files and invoking the device linker, respectively, and link all object files with the intermediate file \texttt{tmp.o}. 
Here, we compare the execution time of the function for tree node in three cases: (1) original implementation, (2) global synchronization using Cooperative Groups, and (3) compiling in the above procedure but executing in the original implementation. 

The measured execution time are \SI{4.0e-3}{s}, \SI{4.9e-3}{s}, and \SI{4.4e-3}{s} for each case, respectively. 
The number of registers used by each thread is \num{56} in the original implementation and \num{64} for the other cases. 
In the latter cases, the number of blocks per SM decreases to \num{8} from \num{9} in the original implementation. 
The reason for the changed compilation method decelerating the application is the decrease of the occupancy with the drop of blocks per SM. 
The additional cost of global synchronization using the Cooperative Groups is estimated to be $\left(4.9 \times 10^{-3} - 4.4 \times 10^{-3}\right) / 21 \simeq 2.3 \times 10^{-5}$~\si{s} because the function performs global synchronization \num{21} times per step.

\bibliographystyle{elsarticle-num}
\newcommand{\apj}{The Astrophysical Journal~}
\newcommand{\apjs}{The Astrophysical Journal Supplement~}
\newcommand{\mnras}{Monthly Notices of the Royal Astronomical Society~}
\newcommand{\na}{New Astronomy~}
\newcommand{\nat}{Nature~}
\newcommand{\aap}{Astronomy and Astrophysics~}
\bibliography{ref}





\end{document}